\ifdefined\pdfminorversion
\fi
\documentclass[sigconf]{acmart}

\usepackage{enumitem}
\usepackage{adjustbox}
\usepackage{wrapfig}
\usepackage{multirow}
\usepackage{graphicx}
\usepackage{array}
\usepackage{amsmath}
\usepackage{booktabs}
\usepackage{placeins}
\usepackage{afterpage}
\usepackage{stfloats}
\usepackage{float}
\usepackage{xcolor}
\usepackage{enumitem}
\usepackage{floatflt}

\usepackage{amssymb,amsfonts}

\makeatletter
\renewcommand{\country}[1]{%
  \global\@ACM@countrypresenttrue
  \ignorespaces
}
\makeatother
\AtBeginDocument{%
  }

\setcopyright{acmlicensed}
\copyrightyear{2026}
\acmYear{2026}
\acmDOI{10.1145/3767308.3835419}
\acmConference[MM '26]{Proceedings of the 34th ACM International Conference on Multimedia}{November 10--13, 2026}{Rio de Janeiro, Brazil}
\acmISBN{979-8-4007-2213-4/2026/11}

\begin{document}

\title{CARA: Concept-Aware Risk Attention for Interpretable Collision Anticipation}

\author{Zhishan Tao\textsuperscript{*}}
\affiliation{%
  \institution{Shanghai Jiao Tong University}
  \country{China}
}
\email{zhishantao@sjtu.edu.cn}

\author{Ruoyu Wang}
\affiliation{%
  \institution{The University of Hong Kong}
  \country{China}
}

\author{Yucheng Wu}
\affiliation{%
  \institution{Fudan University}
  \country{China}
}

\author{Enjun Du}
\affiliation{%
  \institution{The University of Hong Kong}
  \country{China}
}
\affiliation{%
  \institution{The Hong Kong University of Science and Technology (Guangzhou)}
  \country{China}
}

\author{Yilei Yuan}
\affiliation{%
  \institution{Shanghai Jiao Tong University}
  \country{China}
}

\author{Sherwin Ho}
\affiliation{%
  \institution{University of Pennsylvania}
  \country{United States}
}

\author{Yue Su}
\affiliation{%
  \institution{The University of Hong Kong}
  \country{China}
}

\author{Jinbo Su}
\affiliation{%
  \institution{Renmin University of China}
  \country{China}
}

\author{Yi Hong\textsuperscript{\textdagger}}
\affiliation{%
  \institution{Shanghai Jiao Tong University}
  \country{China}
}

\renewcommand{\shortauthors}{Tao et al.}

\begin{abstract}
Collision anticipation in autonomous driving requires not only accurate early warnings but also interpretable reasoning about what risk factors are being tracked and how risk evolves over time. Existing methods fall short in this regard: feature-driven models are opaque, post-hoc explanations often lack fidelity, and concept-based methods are mostly designed for static recognition rather than dynamic driving scenes. We propose \textbf{CARA} (\textbf{C}oncept-\textbf{A}ware \textbf{R}isk \textbf{A}ttention), an intrinsically interpretable spatio-temporal framework for collision anticipation. CARA derives domain-grounded risk concepts from accident narratives, aligns them with video frames via vision--language similarity, and organizes them into evolving concept trajectories. These trajectories provide explicit risk evidence that guides spatial attention, temporal attention, and anticipation, allowing semantic concepts to directly influence both where the model attends and how it predicts risk over time. By treating semantic risk factors as dynamic intermediate evidence rather than auxiliary post-hoc explanations, CARA tightly couples interpretability with the predictive process. Extensive experiments on three benchmarks show that CARA consistently improves anticipation accuracy and warning earliness over strong baselines, while providing sparse and semantically grounded concept evidence.
\end{abstract}




\maketitle

\section{Introduction}


Collision anticipation is a core capability for autonomous driving, yet in safety-critical settings, predictive accuracy alone is insufficient. A trustworthy model should not only forecast an impending collision, but also reveal \emph{what risk factors it is tracking} and \emph{why} risk escalates over time. Without such transparent and auditable reasoning, collision predictors remain difficult to trust, diagnose, and validate in real-world autonomous driving systems \citep{koopman2016challenges, ryan2020artificial, hu2024end, atakishiyev2024explainable}.

\begin{figure}[t]
  \centering
  \includegraphics[width=\columnwidth]{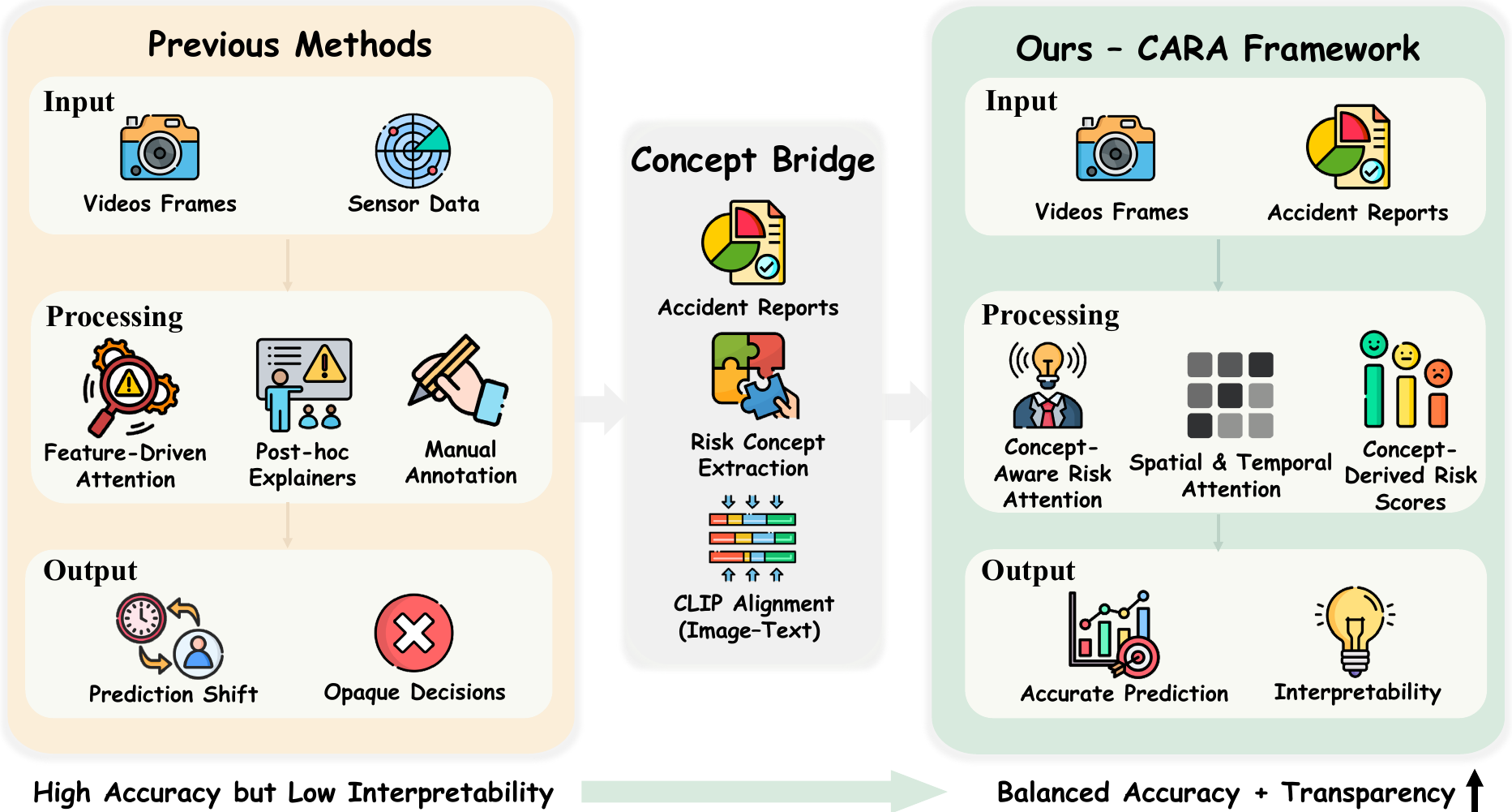}
  \caption{Architectural comparison. (a) Conventional feature-driven paradigms rely on latent spatio-temporal correlations with limited transparency; (b) \textbf{CARA} derives domain-grounded risk concepts and uses them to govern spatial-temporal attention for interpretable collision anticipation.}
  \vspace{-0.34em}
\Description{Comparison between conventional feature-driven collision anticipation and the proposed CARA framework. Conventional methods rely on latent spatio-temporal correlations with limited transparency. CARA derives domain-grounded risk concepts and uses them to guide spatial and temporal attention for interpretable collision anticipation.}
  \label{fig:framework-comparison}
\end{figure}

Despite rapid progress in video-based accident anticipation, most existing methods remain fundamentally feature-driven: they learn spatio-temporal correlations that improve prediction, yet provide limited insight into the underlying risk assessment process \citep{chan2017anticipating, bao2020uncertainty, karim2022dynamic, thakur2024graphgraph, liao2024accnet}. Recent works further incorporate multimodal or language-enhanced context for accident anticipation and localization \citep{liao2024w3al, zhang2025camera}, but still rely primarily on latent feature or modality fusion rather than explicit concept-mediated control. Prior efforts to improve transparency have explored post-hoc explanations and concept-based modeling \citep{ribeiro2016should, arrieta2020explainable, koh2020concept, yuksekgonul2023post, oikarinen2023label}, yet these approaches still fall short for collision anticipation. Post-hoc explanations may highlight influential regions after prediction, but do not guarantee that the explanation faithfully reflects the actual predictive mechanism. Concept-based approaches are more semantically meaningful, yet most are designed for static recognition settings and treat concepts as passive descriptors rather than active variables in sequential decision-making. As a result, current approaches still struggle to provide \emph{faithful} risk reasoning for temporally evolving driving scenes.

In realistic traffic scenarios, collision risk rarely emerges from isolated low-level features alone. Rather, it is shaped by evolving semantic risk factors such as \emph{unsafe merging}, \emph{vulnerable road-user proximity}, \emph{visibility conflict}, or \emph{lane intrusion} \citep{park2021scenario}. These factors form an interpretable evidence stream that unfolds over time and should directly influence both \emph{where} a model attends and \emph{when} it raises risk. This suggests a different formulation of interpretable collision anticipation: instead of predicting first and explaining afterward, a predictor should explicitly track high-level risk concepts as dynamic intermediate signals that govern spatio-temporal reasoning throughout the anticipation process.

A practical challenge, however, is how to obtain such domain-grounded concepts at scale. Manual concept annotation for driving videos is expensive, subjective, and difficult to standardize across diverse accident scenarios. In contrast, accident reports provide a natural source of semantic risk knowledge. Written by humans to describe causal events and hazardous interactions, these reports encode recurring patterns of traffic risk in natural language \citep{park2021scenario, kim2024comprehensive}. This makes them a promising resource for deriving interpretable risk concepts without requiring dense concept labels on video data. Meanwhile, recent advances in vision--language alignment and open-vocabulary grounding make it increasingly feasible to connect language-derived semantics with visual observations \citep{radford2021learning, li2022blip, jia2021scaling}.


Building on these advances, we propose \textbf{Concept-Aware Risk Attention (CARA)}, a concept-grounded interpretable framework for collision anticipation. CARA derives domain-relevant risk concepts from accident narratives and aligns them with video frames via vision--language similarity to obtain frame-level concept evidence. This evidence is then organized into temporally coherent concept trajectories, which serve as explicit risk cues to guide spatial attention, temporal attention, and temporal fusion. In this way, CARA integrates semantic concepts directly into the predictive pathway, enabling both \textbf{semantic-to-spatial} and \textbf{semantic-to-temporal} guidance for anticipatory reasoning. By treating semantic risk factors as dynamic control signals rather than auxiliary explanations, CARA links concept-guided prediction with built-in interpretability. Figure~\ref{fig:framework-comparison} contrasts this concept-mediated formulation with conventional feature-driven paradigms and illustrates how CARA integrates domain-grounded risk concepts into the anticipatory reasoning pipeline.

Experiments on standard benchmarks show that our design consistently improves anticipation accuracy and warning earliness, while producing concept-grounded evidence for prediction. The advantage is particularly evident on the challenging DAD benchmark, where CARA outperforms a strong feature-driven baseline in both average precision and mean warning time. 


Our contributions are summarized as follows:
\begin{itemize}
    \item We formulate interpretable collision anticipation as a \textbf{concept-mediated sequential reasoning} problem, in which semantic risk factors directly participate in prediction rather than serving only as post-hoc explanations.
    
    
    \item We introduce \textbf{CARA}, which derives risk concepts from accident reports, organizes them into temporally evolving concept trajectories, and integrates them into spatial-temporal attention and temporal fusion for interpretable collision anticipation.

    
    \item We show on standard benchmarks that concept-mediated anticipation jointly improves anticipation performance and concept-grounded interpretability, yielding earlier warnings and stronger diagnostic consistency than feature-driven baselines.
\end{itemize}

\section{Related Work}

\subsection{Traffic Accident Anticipation Models}

Traffic accident anticipation has advanced through recurrent, graph-based, and Transformer architectures \citep{chan2017anticipating, bao2020uncertainty, karim2022dynamic, thakur2024graphgraph}. Additional modalities and structured cues, including depth, text, and language-enhanced localization, further improve prediction \citep{shao2024lmdrive, mao2023gpt, liao2024accnet, zhang2025camera, liao2024w3al}. Beyond driving, multi-view contrastive learning integrates complementary relational signals and mitigates noisy supervision in graph-structured domains \citep{Yang2025DSVC}. Text-enhanced anticipation similarly exploits accident reports or structured priors \citep{guan2025domain}. Yet these approaches remain largely feature-driven, offering limited visibility into the semantic evidence behind evolving risk. CARA instead grounds prediction in explicit risk factors represented as traceable concept trajectories.


\vspace{-0.2em}
\subsection{Explainable and Concept-Based Approaches}

Explainability research spans post-hoc, language, concept-based, and autonomous-driving approaches \citep{zablocki2022explainability}. Saliency and attribution methods highlight influential regions but may not faithfully reflect the decision process \citep{selvaraju2017grad, lundberg2017unified, ribeiro2016should, yeh2019infidelity}, motivating intrinsically interpretable models for high-stakes settings \citep{rudin2019stop}. Language-based explanations describe driving behavior \citep{kim2018textual, kim2020advisable}, yet are typically generated after prediction and weakly coupled to it. Counterfactual trajectory analysis and query-based motion planning \citep{hsu2023counterfactual, biswas2024quad} improve interpretability, but target forecasting or planning rather than collision anticipation with explicit semantic control.

Concept Bottleneck Models (CBMs) expose intermediate semantic variables \citep{koh2020concept, yuksekgonul2023post, oikarinen2023label, yang2023language}, but usually target static recognition with time-invariant concepts. NLP and LLM reasoning indicates that textual knowledge can provide scalable priors without dense manual labels \citep{park2021scenario, kim2024comprehensive, du2025graphmaster, du2026graphoracle, du2025mokgr}. Collision anticipation still lacks a model in which semantic concepts evolve over time and directly govern sequential prediction, rather than serving as auxiliary labels or post-hoc descriptors. CARA addresses this gap by converting concepts from accident narratives into temporal trajectories that modulate spatio-temporal attention throughout anticipation.

\begin{figure*}[t]
  \centering
  \includegraphics[width=0.97\textwidth]{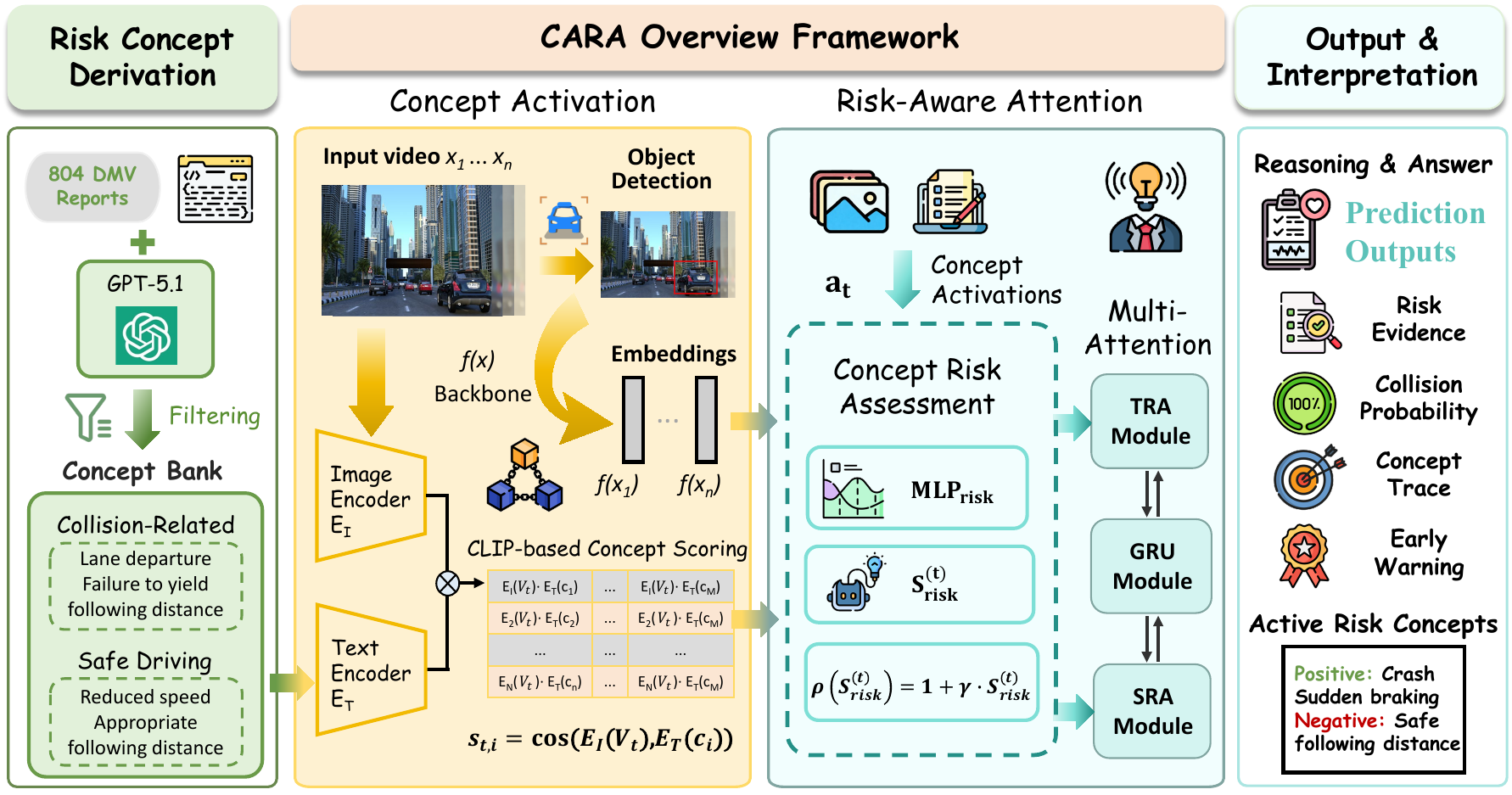}
  \caption{\textbf{CARA Framework Overview.} CARA derives risk-aware concepts from 804 DMV accident reports and aligns them with video observations using CLIP. These concept signals are organized into temporal trajectories and used for concept risk assessment, which in turn guides spatial and temporal attention for interpretable collision anticipation.
  }
  \vspace{-0.3em}
  \label{fig:framework}
\end{figure*}

\vspace{-0.2em}
\section{Methodology}

\subsection{Problem Formulation}

Traffic collision anticipation aims to estimate frame-wise collision probability and provide early warnings before an accident occurs.
Given a dashboard video stream of $T$ frames $\mathcal{V}=\{V_1,V_2,\ldots,V_T\}$, the task is to predict, at each time step, frame-wise collision probabilities $\mathcal{P}=\{p_1,p_2,\ldots,p_T\}$ together with a sequence of concept activation vectors $\mathcal{A}=\{\mathbf{a}_1,\mathbf{a}_2,\ldots,\mathbf{a}_T\}$, where $\mathbf{a}_t=[a_{t,1},a_{t,2},\ldots,a_{t,K}]$ and $K$ denotes the size of the interpretable concept library.

For accident-positive videos with collision time $\tau$, we define the Time-to-Accident (TTA) as $\Delta t=\tau-t_o$, where $t_o=\min\{t\mid p_t\ge p_o\}$ denotes the earliest frame at which the predicted probability exceeds a threshold $p_o$. A video is classified as accident-positive if $\exists\, t\le \tau$ such that $p_t\ge p_o$, and accident-negative if $\tau=0$. This formulation emphasizes three key requirements: accurate probability estimation, early warning characterized by a large $\Delta t$, and interpretability through human-understandable semantic states. In CARA, the concept activation sequence $\{\mathbf{a}_1,\ldots,\mathbf{a}_T\}$ is not merely an auxiliary explanatory output, but an explicit semantic state that supports sequential risk reasoning throughout prediction.

\vspace{-0.6em}
\subsection{Overview of CARA Framework}

As illustrated in Fig.~\ref{fig:framework}, CARA anticipates collisions through concept-mediated sequential reasoning. Rather than directly predicting risk from latent spatio-temporal correlations, it grounds semantic risk concepts, converts them into dynamic attention guidance, and preserves them during temporal prediction.

Concretely, CARA consists of three stages: (1) \textit{Risk Concept Derivation and Grounding} (Sec.~\ref{sec:concept_generation}), which derives domain-grounded concepts from accident reports and grounds them into frame-wise semantic activations; (2) \textit{Concept-Guided Risk Attention} (Sec.~\ref{sec:cara}), which transforms temporally evolving concept trajectories into risk evidence that guides spatial and temporal attention; and (3) \textit{Concept-Preserving Temporal Prediction} (Sec.~\ref{sec:fusion}), which preserves concept semantics throughout sequential prediction.

\subsection{Stage I: Risk Concept Derivation and Grounding}
\label{sec:concept_generation}

The first stage constructs a domain-grounded concept space and maps it to frame-wise semantic evidence.

\vspace{+0.3em}
\noindent \textbf{Concept Derivation from Accident Narratives.}
We use 804 California DMV autonomous vehicle accident reports (January 2019 to March 2025) to derive domain-grounded traffic risk concepts. Using spaCy dependency parsing, we extract collision-related noun phrases, vehicle behaviors, causal relations, and environmental factors, yielding approximately 1,840 candidate concepts. To balance risk and non-risk semantics, we leverage GPT-5.1 to generate context-preserving safe-driving descriptions by replacing risky behaviors with safer alternatives. After deduplication, frequency filtering, and relevance filtering, the candidate set is reduced to 892 concepts. We then retain only concepts with sufficient visual grounding on held-out driving frames under a CLIP-based criterion, resulting in a final library of 210 interpretable concepts spanning vehicle behaviors, environmental conditions, road-user interactions, traffic violations, and safe-driving cues. Full extraction rules, filtering criteria, and concept statistics are provided in Appendix~B.


\vspace{+0.3em}
\noindent \textbf{CLIP-based Concept Scoring and Temporal Stabilization.}
For each video frame $V_t$, we compute a CLIP image embedding $E_I(V_t)\in\mathbb{R}^d$ and encode each concept description $c_i$ with a CLIP text embedding $E_T(c_i)\in\mathbb{R}^d$, where $\mathcal{C}=\{c_1,\ldots,c_K\}$. The raw concept score is defined as the cosine similarity
\begin{equation}
s_{t,i}=\cos\!\left(E_I(V_t),\,E_T(c_i)\right).
\end{equation}
Here, the CLIP encoders are frozen to preserve a stable semantic space. A lightweight learnable calibration module then adapts the resulting generic similarities to collision-oriented concept channels:
\begin{equation}
\hat{s}_{t,i}=\sigma\!\left(w_i \cdot s_{t,i}\right),
\end{equation}
where $w_i$ is a learnable per-concept scalar, and $\hat{s}_{t,i}$ denotes the calibrated concept score for concept $c_i$ at frame $t$. To reduce frame-level instability, we apply exponential moving average (EMA) smoothing:
\begin{equation}
a_{t,i}=\eta\cdot \hat{s}_{t,i}+(1-\eta)\cdot a_{t-1,i}.
\end{equation}
where $\eta$ is the smoothing factor. We initialize $a_{1,i}=\hat{s}_{1,i}$ and apply the recursion for $t\ge2$, yielding $\mathbf{a}_t\in\mathbb{R}^{K}$ as a stable semantic trace (Appendix~F.2). EMA transforms isolated frame-level concept matches into temporally coherent trajectories, making downstream reasoning depend on persistent semantic evidence rather than transient noise. Importantly, CLIP image embeddings are used only to compute concept scores as semantic anchors, and are not used as predictive visual features for collision anticipation.

\subsection{Stage II: Concept-Guided Risk Attention}
\label{sec:cara}

The second stage converts concept trajectories into risk evidence (i.e., risk components)  for \textbf{Spatial Risk Attention (SRA)} and \textbf{Temporal Risk Attention (TRA)},  which together form risk-aware attention.

\vspace{+0.3em}
\noindent \textbf{Concept-Mediated Risk Assessment.}
Traditional models (e.g., DSTA~\cite{karim2022dynamic}) learn attention weights end-to-end from collision supervision, often lacking interpretable grounding. In contrast, CARA introduces a concept-mediated hierarchy in which the concept activation vector $\mathbf{a}_t$ from Sec.~\ref{sec:concept_generation} explicitly determines both spatial and temporal attention:
\begin{equation}
\bigl(\boldsymbol{\alpha}_t^{\text{spatial}},\boldsymbol{\beta}_t^{\text{temporal}}\bigr)=f_\theta(\mathbf{a}_t).
\end{equation}
Here, $f_\theta(\cdot)$ denotes the concept-to-attention mapping from concept activations to spatial and temporal attention guidance. This design yields intrinsic interpretability without relying on post-hoc attribution methods such as GradCAM~\cite{selvaraju2017grad} or SHAP~\cite{lundberg2017unified}.

Temporally stabilized concept trajectories become useful for anticipation when they are aggregated into risk evidence. Given $\mathbf{a}_t$, the \textbf{Concept Risk Assessment (CRA)} module aggregates concept-level risk components into a frame-level risk score:
\begin{equation}
S_{\text{risk}}^{(t)}=\sigma\!\left(\text{MLP}_{\text{risk}}(\mathbf{a}_t)\right),
\end{equation}
where $\sigma$ is sigmoid and $\text{MLP}_{\text{risk}}$ learns to weight concept activations by collision relevance. This risk score acts as a contextual gate that determines how strongly current semantic evidence should influence subsequent spatial and temporal attention. We define the risk modulation function as
\begin{equation}
\rho\!\left(S_{\text{risk}}^{(t)}\right)=1+\gamma\cdot S_{\text{risk}}^{(t)},
\label{eq:risk}
\end{equation}
where $\gamma$ is the amplification factor. The modulator $\rho(\cdot)$ therefore amplifies attention under higher concept-derived risk.

\vspace{+0.3em}
\noindent \textbf{Risk-Guided Spatial and Temporal Attention.}
For \textbf{Spatial Risk Attention (SRA)}, the standard attention logits
$\mathbf{e}_t^{\text{spatial}} = [e_{t,1}^{\text{spatial}}, \ldots, e_{t,N}^{\text{spatial}}]$ are
computed from object-level features $x_{t,i}$ and recurrent state $h_{t-1}$. Object features are extracted from the top-$N$ detector proposals in each frame and encoded with a VGG-16 backbone, following common object-centric anticipation practice (see Appendix~C.3 for extraction details). To enable concept conditioned \emph{object-wise} reallocation, we introduce a concept-conditioned object bias vector
$\mathbf{b}_t=[b_{t,1},\ldots,b_{t,N}]\in\mathbb{R}^{N}$:
\begin{equation}
b_{t,i}=\text{MLP}_{\text{obj}}\!\left(\mathbf{x}_{t,i}\,\|\,\mathbf{a}_t\right),
\label{eq:object_bias}
\end{equation}
where $\|$ denotes concatenation. The risk modulator then scales this bias and yields the modulated spatial attention:
\begin{equation}
\boldsymbol{\alpha}_t^{\text{spatial}}
=\text{softmax}\!\left(\mathbf{e}_t^{\text{spatial}}
+\rho\!\left(S_{\text{risk}}^{(t)}\right)\cdot \mathbf{b}_t\right),
\label{eq:spatial_attention}
\end{equation}

\vspace{+0.3em}
For \textbf{Temporal Risk Attention (TRA)} over a sliding window of $M$ frames, we define
\begin{equation}
\resizebox{0.75\columnwidth}{!}{$\displaystyle
\boldsymbol{\beta}_t^{\text{temporal}}=\text{softmax}\!\left(
\mathbf{e}_t^{\text{temporal}}\odot\phi\!\left(S_{\text{risk}}^{(t-M+1:t)}\right)
\right),
$}
\end{equation}
where $\phi(\cdot)$ is a 1D causal convolution that captures temporal risk patterns using past and current frames (Appendix~C.5). Spatial attention localizes hazard-relevant objects, while temporal attention emphasizes sustained risk escalation over transient fluctuations. Together, the two branches model both local hazard concentration and the temporal accumulation of collision evidence.

\subsection{Stage III: Concept-Preserving Temporal Prediction}
\label{sec:fusion}

The final stage performs temporal prediction with a \textbf{GRU-based} fusion module. Let $\mathbf{h}_t$ denote the GRU hidden state, which serves as the concept-preserving temporal representation for final prediction. Merely modulating attention is insufficient if semantic evidence is not retained during temporal accumulation, since recurrent fusion may dilute or drift away from high-level concept signals. To prevent this, the GRU injects the concept activation vector $\mathbf{a}_t$ at every step:
\begin{equation}
\mathbf{h}_t=\text{GRU}(\mathbf{f}_t^{\text{attended}}\| \mathbf{a}_t,\mathbf{h}_{t-1}),
\end{equation}
where $\mathbf{f}_t^{\text{attended}}$ is the concept-aware spatio-temporally weighted visual feature and $\|$ denotes concatenation. By explicitly fusing attended visual cues with concept-derived semantic anchors at each time step, CARA preserves semantic fidelity during temporal accumulation and keeps prediction conditioned on explicit concept trajectories. The prediction head $g_\phi$ then maps $\mathbf{h}_t$ to $p_t$. Although $\mathbf{h}_t$ is not directly interpretable, $\{\mathbf{a}_1,\ldots,\mathbf{a}_T\}$ remains an accessible semantic trace of risk assessment (Appendix~D).

\subsection{Training Objective}
\label{sec:training_loss}

To jointly optimize predictive performance and concept-level interpretability, we train CARA with the following multi-task objective:
\begin{equation}
\mathcal{L}_{\text{total}}
=
\mathcal{L}_{\text{collision}}
+\lambda_1 \mathcal{L}_{\text{concept}}
+\lambda_2 \mathcal{L}_{\text{interpretability}}.
\end{equation}
The collision term $\mathcal{L}_{\text{collision}}$ optimizes frame-wise anticipation accuracy, the concept consistency term $\mathcal{L}_{\text{concept}}$ prevents concept channels from drifting away from their CLIP-based semantic anchors, and the interpretability term $\mathcal{L}_{\text{interpretability}}$ encourages compact and selective concept usage. This objective formulation ensures that CARA learns not only when to warn, but also which semantic evidence should remain active and interpretable throughout temporal reasoning. Together, these objectives help CARA produce accurate early warnings while preserving semantically meaningful and concise concept evidence. For brevity, we summarize their roles here and defer the full loss definitions and extended analysis to Appendix~E, Appendix~F.3, and Appendix~A.3.

\setlength{\tabcolsep}{5pt}
\begin{table*}[t] 
\centering
\caption{Performance comparison on three benchmarks. \textbf{Bold}: the best, \underline{underline}: the second-best.}
\small
\begin{tabular}{l|ccc|ccc|ccc}
\toprule
\multirow{2}{*}{\textbf{Model}} & \multicolumn{3}{c|}{\textbf{DAD}} & \multicolumn{3}{c|}{\textbf{A3D}} & \multicolumn{3}{c}{\textbf{CCD}} \\
& \textbf{AP(\%)} & \textbf{mTTA(s)} & \textbf{R80(s)} & \textbf{AP(\%)} & \textbf{mTTA(s)} & \textbf{R80(s)} & \textbf{AP(\%)} & \textbf{mTTA(s)} & \textbf{R80(s)} \\
\midrule
DSA~\cite{chan2017anticipating} & 63.37 & 1.58 & 1.81 & 93.58 & 3.61 & 4.13 & 98.10 & 3.97 & 4.21 \\
UString~\cite{bao2020uncertainty} & 68.10 & 1.51 & 2.13 & 94.08 & 3.96 & 4.61 & 98.53 & 4.55 & 4.82 \\
DSTA~\cite{karim2022dynamic} & 66.69 & 1.61 & \underline{2.23} & 93.71 & 3.87 & 4.67 & 98.67 & 4.33 & 4.59 \\
GSC~\cite{wang2023gsc} & 68.70 & 1.29 & 2.11 & 93.89 & 3.76 & 4.51 & 98.95 & 4.29 & 4.57 \\
CRASH~\cite{crash} & \underline{70.51} & \underline{1.78} & 2.16 & \underline{94.17} & \underline{4.61} & \underline{4.89} & \underline{99.13} & \underline{4.63} & \underline{4.83} \\
\textbf{CARA (Ours)} & \textbf{75.37} & \textbf{1.97} & \textbf{2.27} & \textbf{95.63} & \textbf{4.72} & \textbf{4.93} & \textbf{99.35} & \textbf{4.79} & \textbf{4.87} \\
\bottomrule
\end{tabular}

\label{tab:main_results}
\end{table*}

\section{Experiments}

We conduct experiments to answer four research questions:
\textbf{RQ1:} Does CARA outperform state-of-the-art methods in accuracy and early warning (Sec.~\ref{subsec:main_results})?
\textbf{RQ2:} How essential are CARA’s risk-aware components and its native integration of semantic concepts (Sec.~\ref{subsec:ablation})?
\textbf{RQ3:} Does CARA provide interpretable, faithful, and decision-relevant concept-based explanations (Sec.~\ref{subsec:interpretability})?
\textbf{RQ4:} Do CARA’s gains depend on meaningful risk concept semantics and temporally coherent concept trajectories, rather than incidental structural artifacts (Sec.~\ref{subsec:rq4_sanity_temporal})?

\subsection{Experimental Setup}
\label{sec:setup}

CARA is evaluated on three collision anticipation benchmarks: \textbf{DAD}~\citep{chan2017anticipating}, \textbf{CCD}~\citep{bao2020uncertainty}, and \textbf{A3D}~\citep{yao2019unsupervised}. Following prior work~\citep{chan2017anticipating, karim2022dynamic}, we report \textbf{Average Precision (AP)}, \textbf{mean Time-to-Accident (mTTA)}, and \textbf{TTA@R80 (R80)} to measure both discrimination and warning earliness. We compare against five representative baselines: \textbf{DSA}~\citep{chan2017anticipating}, \textbf{UString}~\citep{bao2020uncertainty}, \textbf{DSTA}~\citep{karim2022dynamic}, \textbf{GSC}~\citep{wang2023gsc}, and \textbf{CRASH}~\citep{crash}. CARA uses 210 risk concepts derived from real-world accident reports and embedded by a frozen CLIP ViT-B/32 encoder~\citep{radford2021learning}; \textbf{these concepts act as intermediate semantic signals rather than additional predictive visual features}. Further details and semantic-anchor analyses are provided in Appendices~A and~G.5.


\subsection{Overall Performance (RQ1)}
\label{subsec:main_results}

\noindent
\textbf{To answer RQ1}, Table~\ref{tab:main_results} compares CARA with state-of-the-art methods across the three benchmarks. \textbf{CARA consistently improves both prediction accuracy and warning earliness.}


On the challenging \textbf{DAD} dataset, CARA improves AP from 70.51\% to 75.37\% and mTTA from 1.78s to 1.97s over the strongest baseline, CRASH, while also achieving the best R80. The simultaneous improvement in accuracy and both early-warning metrics is important because it rules out a trivial trade-off in which earlier alarms are obtained by simply lowering the decision threshold. \textbf{Instead, CARA better discriminates temporally ambiguous interactions while raising risk earlier.}

\begin{table}[t]
\centering
\setlength{\tabcolsep}{3pt}
\small
\caption{Apples-to-apples comparison with concept bottleneck models on the DAD dataset.}
\begin{tabular}{l|ccc}
\toprule
\textbf{Model} & \textbf{AP(\%)} & \textbf{mTTA(s)} & \textbf{R80(s)} \\
\midrule
Baseline (DSTA) & 66.69 & 1.61 & 2.23 \\
CBM (Concept-to-Prediction) & 64.80 & 1.30 & 2.05 \\
CBM + Temporal Aggregation & 66.80 & 1.58 & 2.20 \\
\textbf{CARA (Ours)} & \textbf{75.37} & \textbf{1.97} & \textbf{2.27} \\
\bottomrule
\end{tabular}
\label{tab:cbm_comparison}
\end{table}
\vspace{-0.3em}

Consistent gains on \textbf{A3D} and \textbf{CCD} show that the benefit is not confined to the visual statistics of DAD. This cross-dataset behavior suggests that the intermediate semantic layer captures transferable risk structure across different driving distributions. The advantage also remains stable across three random seeds with low variance relative to CRASH (Appendix~G). \textbf{Overall, CARA improves both collision discrimination and the timeliness of anticipation by grounding prediction in semantically meaningful risk evidence.}

\begin{table*}[t]
\centering
\begin{minipage}[t]{0.48\textwidth}
\centering
\small
\caption{CBM integration analysis: Performance drop when adding CBM to existing methods vs. CARA's native concept design.}
\adjustbox{max width=\linewidth}{
\begin{tabular}{l|ccc}
\toprule
\textbf{Model} & \textbf{DAD AP(\%)} & \textbf{A3D AP(\%)} & \textbf{CCD AP(\%)} \\
\midrule
UString & 68.10 & 94.08 & 98.53 \\
UString+CBM & 65.80 ($\downarrow$2.30) & 92.50 ($\downarrow$1.58) & 97.80 ($\downarrow$0.73) \\
\hline
DSTA & 66.69 & 93.71 & 98.67 \\
DSTA+CBM & 64.10 ($\downarrow$2.59) & 92.00 ($\downarrow$1.71) & 98.00 ($\downarrow$0.67) \\
\hline
CRASH & 70.51 & 94.17 & 99.13 \\
CRASH+CBM & 68.90 ($\downarrow$1.61) & 93.10 ($\downarrow$1.07) & 98.60 ($\downarrow$0.53) \\
\hline
\textbf{CARA (Ours)} & \textbf{75.37} & \textbf{95.63} & \textbf{99.35} \\
\bottomrule
\end{tabular}
}
\label{tab:cbm_ablation}
\end{minipage}
\hfill
\begin{minipage}[t]{0.48\textwidth}
\centering
\small
\caption{Component ablation on the DAD dataset. CRA: Concept Risk Assessment, SRA: Spatial Risk Attention, TRA: Temporal Risk Attention.}
\adjustbox{max width=\linewidth}{
\begin{tabular}{l|ccc}
\toprule
\textbf{Variant} & \textbf{AP(\%)} & \textbf{mTTA(s)} & \textbf{R80(s)} \\
\midrule
\textbf{Full CARA} & \textbf{75.37} & \textbf{1.97} & \textbf{2.27} \\
\hline
w/o CRA & 68.92 ($\downarrow$6.45) & 1.65 ($\downarrow$0.32) & 1.98 ($\downarrow$0.29) \\
w/o SRA & 69.75 ($\downarrow$5.62) & 1.75 ($\downarrow$0.22) & 2.08 ($\downarrow$0.19) \\
w/o TRA & 69.25 ($\downarrow$6.12) & 1.63 ($\downarrow$0.34) & 1.97 ($\downarrow$0.30) \\
w/o $\mathcal{L}_{\text{concept}}$ & 70.05 ($\downarrow$5.32) & 1.83 ($\downarrow$0.14) & 2.10 ($\downarrow$0.17) \\
w/o $\mathcal{L}_{\text{interpretability}}$ & 70.32 ($\downarrow$5.05) & 1.84 ($\downarrow$0.13) & 2.10 ($\downarrow$0.17) \\
\hline
w/o Risk-Aware Attn & 68.15 ($\downarrow$7.22) & 1.42 ($\downarrow$0.55) & 1.72 ($\downarrow$0.55) \\
w/o Risk Components & 67.05 ($\downarrow$8.32) & 1.05 ($\downarrow$0.92) & 1.42 ($\downarrow$0.85) \\
\bottomrule
\end{tabular}
}
\label{tab:component_ablation}
\end{minipage}
\end{table*}

\vspace{+0.2em}
\subsection{Ablation Study on CARA (RQ2)}
\label{subsec:ablation}

\noindent \textbf{To address RQ2}, we compare CARA's native concept integration with post-hoc CBM retrofitting and examine how its risk-aware components and auxiliary objectives contribute to accuracy, early warning, and semantic stability.

\begin{figure*}[t]
  \vspace{-0.3em}
  \centering

  \begin{minipage}[t]{0.48\textwidth}
    \centering
    \includegraphics[width=0.85\linewidth]{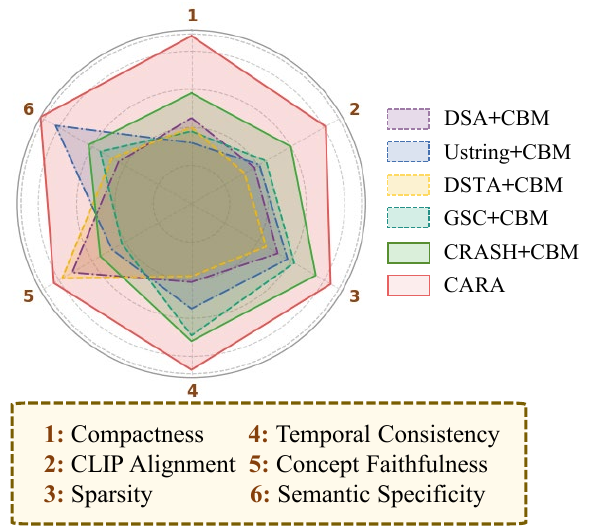}

    \makeatletter
    \def\@captype{figure}
    \makeatother
   \caption{Comparison of holistic concept quality between CARA and post-hoc CBM baselines across six interpretability dimensions.}
    \label{fig:radar_concept_quality}
  \end{minipage}
  \hfill
  \begin{minipage}[t]{0.48\textwidth}
    \centering
    \includegraphics[width=0.73\linewidth]{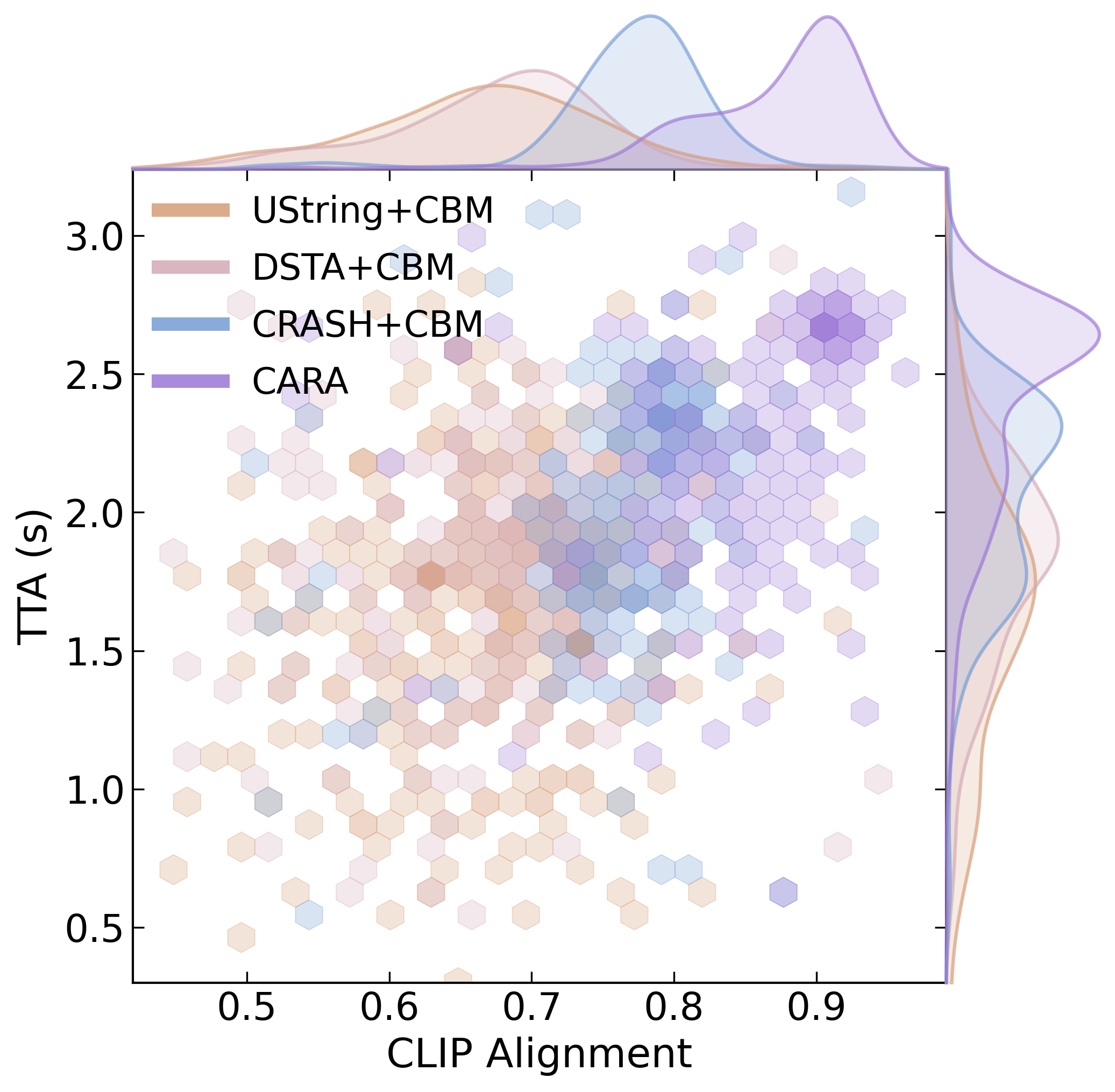}

    \makeatletter
    \def\@captype{figure}
    \makeatother
    \caption{Comparison of the relationship between clip-level CLIP alignment and warning lead time (TTA) for CARA and post-hoc CBM baselines.}
    \label{fig:alignment_tta_joint}
  \end{minipage}

\end{figure*}

\vspace{+0.1em}
\subsubsection{Necessity of Native Concept Integration.}


Table~\ref{tab:cbm_comparison} shows that forcing the same concepts to serve as the sole predictive pathway degrades both accuracy and warning earliness, while temporal aggregation provides only partial recovery. Holding the concept inventory fixed isolates the integration strategy rather than its content. Concept activations can be noisy or temporally immature, making them insufficient as a complete replacement for visual evidence. In contrast, CARA uses concepts to guide risk estimation and attention on top of high-capacity visual representations. \textbf{This design preserves information flow, contextualizes semantic cues with the scene, and allows concepts to shape prediction without becoming a restrictive bottleneck.}

\vspace{+0.1em}

\subsubsection{CBM Integration Analysis.}

Table~\ref{tab:cbm_ablation} further shows that retrofitting CBM modules consistently degrades AP across all datasets. The cross-dataset consistency indicates that this is a structural limitation of post-hoc integration rather than an isolated dataset effect. Such scenes require semantic cues to interact with spatial and temporal representations throughout prediction; an output-side bottleneck instead amplifies semantic--visual misalignment and disrupts end-to-end optimization. CARA jointly learns concept representations with the predictor and preserves their alignment through $\mathcal{L}_{\text{concept}}$, allowing concepts to modulate attention as evidence evolves. \textbf{These findings show that native concept integration, rather than post-hoc attachment, is critical to CARA's gains.}



\subsubsection{Component Analysis.}

Table~\ref{tab:component_ablation} shows complementary roles for CARA's components. \textbf{CRA} converts concept trajectories into risk estimates, improving semantic conditioning of the visual representation; \textbf{SRA} reallocates attention toward concept-consistent hazards; and \textbf{TRA} accumulates persistent evidence to support early warning. The observed degradation pattern is consistent with these intended roles: spatial and temporal modules affect different aspects of anticipation, while CRA links both to semantic risk. Removing any component degrades performance, and jointly removing the risk components causes the largest decline, confirming that semantic conditioning, spatial focus, and temporal accumulation are mutually reinforcing. The auxiliary losses provide a second source of support by maintaining concept consistency and discouraging diffuse activations. \textbf{Similar trends on A3D and CCD further show that the gains arise from coordinated mechanisms rather than a single dominant module.} Proposal sensitivity is reported in Appendix~C.4.

\vspace{+0.1em}

\subsubsection{Computational Efficiency.}

The concept-driven modules add 5--8\% computational overhead relative to CRASH while keeping parameter count and training time comparable. Because the backbone remains comparable, the performance gains cannot be attributed to a substantially larger model or disproportionate training budget. \textbf{Thus, semantic control and interpretability are introduced without substantially increasing computational cost} (Appendix~A).

\vspace{-0.3em}

\subsection{Interpretability Analysis (RQ3)}
\label{subsec:interpretability}

\noindent \textbf{To address RQ3}, we evaluate CARA's interpretability from four aspects:
(i) a holistic summary of concept quality across multiple dimensions,
(ii) the relationship between concept quality and early warning at the clip level,
(iii) concept semantic faithfulness and qualitative decision mediation, and
(iv) semantically traceable failure modes and forensic utility.

\begin{figure*}[t]
  \centering
  \includegraphics[width=0.97\textwidth]{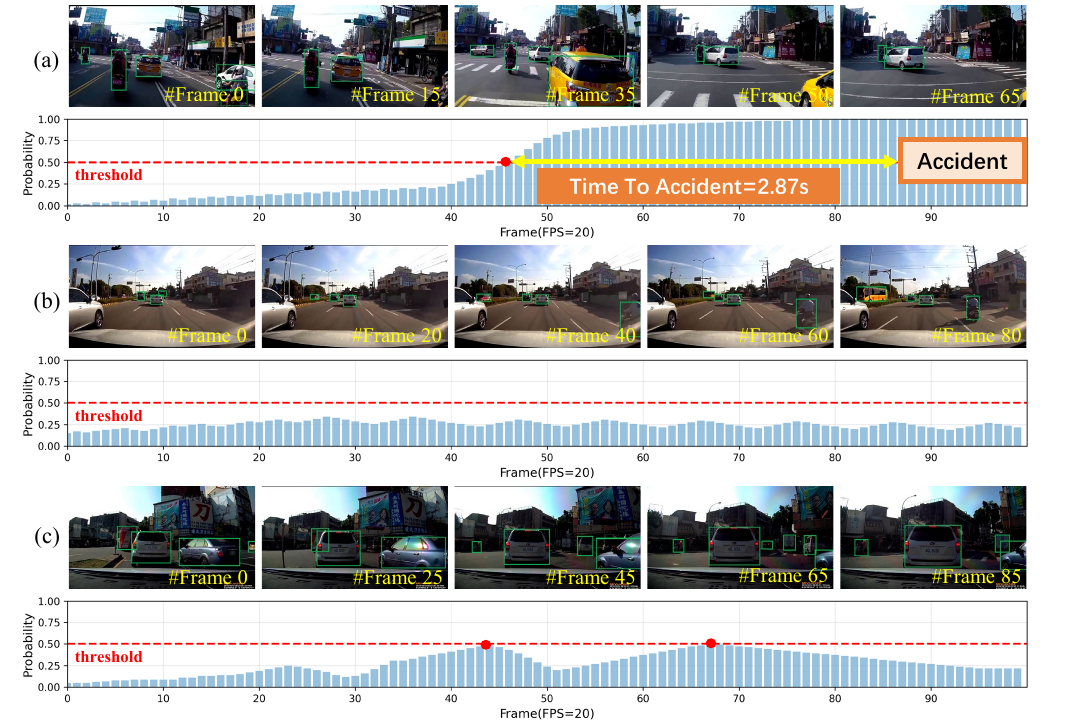}
  \caption{\textbf{Case studies of collision anticipation on DAD.} Frame-wise prediction probabilities (blue bars) and attention visualization. The red dashed line indicates the 0.5 threshold, and green boxes mark top-attended objects. Additional case studies are provided in Appendix~L.}
  \Description{Case studies of collision anticipation on DAD showing frame-wise prediction probabilities, a red dashed threshold at 0.5, and attention visualization with green boxes marking top-attended objects.}
  \label{fig:attention_cases}
\end{figure*}

\subsubsection{Holistic Concept Quality Overview.}
\label{subsec:rq3_holistic}

\noindent
Figure~\ref{fig:radar_concept_quality} compares CARA with post-hoc CBM baselines across six normalized interpretability dimensions: compactness, CLIP alignment, sparsity, temporal consistency, concept faithfulness, and semantic specificity. Joint evaluation avoids reliance on a single proxy. CARA shows the most balanced outward profile, indicating broad improvements rather than optimization for one measure. The clearest gains occur in compactness, faithfulness, and temporal consistency, where post-hoc bottlenecks tend to produce diffuse or unstable concept traces. CARA instead maintains a smaller set of semantically relevant concepts as the scene evolves. \textbf{Its explanations are therefore not only grounded, but also cleaner, more temporally stable, and more behaviorally specific.} Metric definitions and quantitative details are provided in Appendix~I.

\vspace{+0.2em}
\subsubsection{Concept Quality and Early Warning Coupling.}
\label{subsec:rq3_coupling}

\noindent
Figure~\ref{fig:alignment_tta_joint} examines whether concept quality is related to anticipation behavior at the sample level. Clips with better CLIP alignment generally exhibit longer warning lead time, and CARA shifts the joint distribution toward the upper-right region, where both properties are strong. Quantitatively, CARA yields the strongest alignment--TTA correlation (Pearson $r=0.58$ vs. $0.42$ for CRASH+CBM). Although this analysis is correlational, the consistent joint and marginal trends suggest that better semantic grounding supports earlier evidence accumulation rather than serving as a descriptive output after prediction. \textbf{CARA's interpretability is therefore coupled with functional anticipation behavior.} Additional distributional statistics and high-quality-region analyses are provided in Appendix~J.

\vspace{+0.2em}
\subsubsection{Semantic Faithfulness and Decision Mediation.}
\label{subsec:rq3_faithfulness_mediation}

\noindent
We evaluate semantic faithfulness on 150 manually annotated clips, with 50 clips from each dataset and labels for eight representative risk concepts. The annotations record whether each concept is visually present, providing an external reference independent of the learned embedding space. Each target concept is evaluated from its projected and temporally stabilized activation using ROC-AUC. Evaluating these activations reflects the signals actually consumed by the predictor rather than an isolated probe of the original text or image encoder. \textbf{The resulting macro-average AUC of 0.871 demonstrates alignment with observable driving events rather than only embedding-space similarity.} This protocol also tests whether concept channels retain their intended semantics after integration with the predictor. Per-dataset, per-concept, and expanded probing results are reported in Appendix~G.

Figure~\ref{fig:attention_cases} further shows how these concepts mediate prediction. In positive cases, emerging risk concepts increase collision probability and concentrate attention on the corresponding hazards. In negative cases, persistent safety concepts keep risk below the warning threshold; in confusing negatives, they counterbalance transient hazard cues and suppress false alarms. \textbf{Together, the quantitative and qualitative results show that CARA provides semantically grounded, decision-relevant evidence throughout anticipation.} Additional cases are provided in Appendix~L.

\vspace{+0.2em}
\subsubsection{Failure Traceability and Forensic Utility.}
\label{subsec:rq3_failure_traceability}


\noindent Under severe occlusion, abrupt maneuvers, or cluttered interactions, CARA's failures remain semantically traceable. For example, a false negative can be linked to under-activation of a relevant concept caused by occlusion rather than opaque latent drift. This distinction helps separate failures in concept evidence from errors in grounding or temporal stabilization. Unlike an undifferentiated confidence score, the resulting trace indicates whether refinement should target the concept library, visual alignment, or difficult-scene augmentation. \textbf{Thus, CARA's interpretability supports not only explanation but also targeted diagnosis of failure modes in safety-critical settings.}

\vspace{+0.2em}
\subsection{Concept Sanity and Temporal Necessity Analysis (RQ4)}
\label{subsec:rq4_sanity_temporal}

\begin{figure}[t]
  \centering
  \includegraphics[width=\columnwidth]{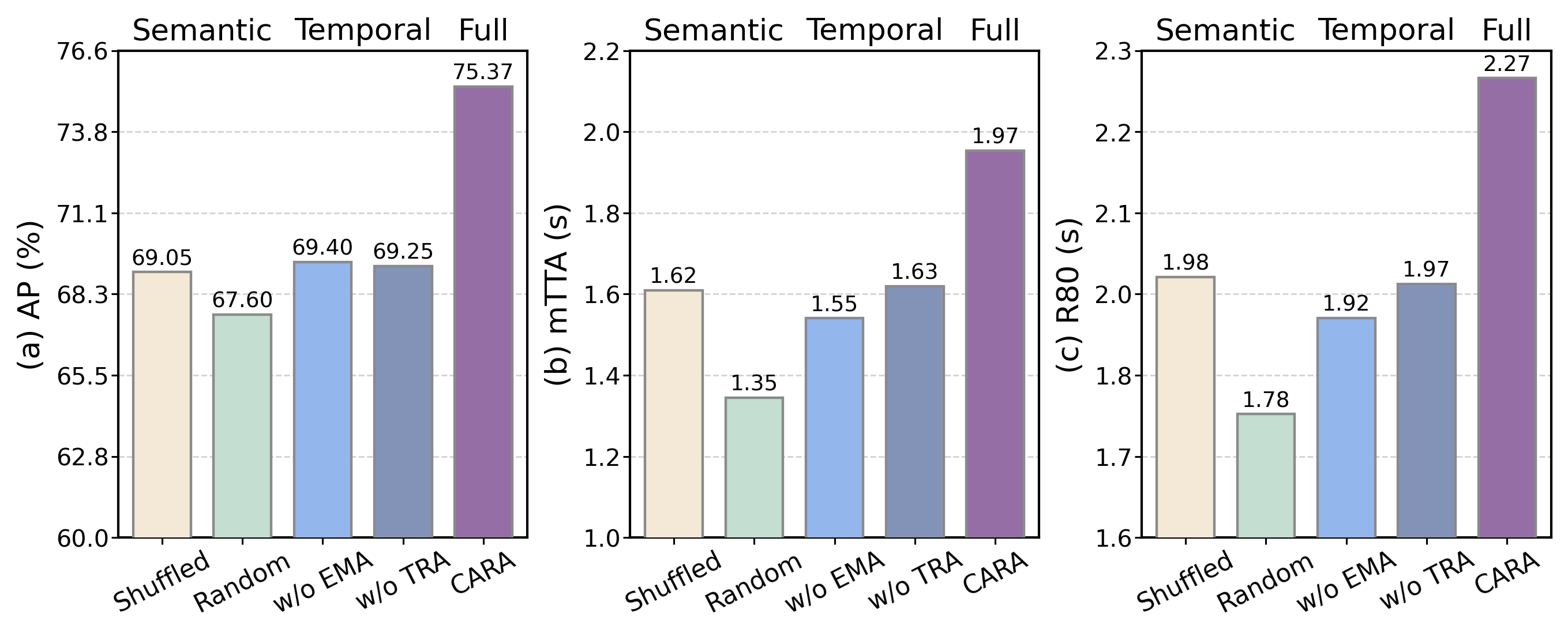}
  \caption{RQ4 analysis on DAD. The three panels visualize AP, mTTA, and R80 for semantic sanity checks (\textit{Shuffled}, \textit{Random}), temporal necessity checks (\textit{w/o EMA}, \textit{w/o TRA}), and the full CARA model.}
  \vspace{-0.7em}
  \label{fig:rq4_counterfactual}
\end{figure}

\noindent \textbf{To address RQ4}, we perturb concept semantics and temporal mechanisms on DAD while preserving the overall model structure. These controlled interventions isolate whether performance depends on meaningful semantic identity and coherent temporal evolution. \textbf{Figure~\ref{fig:rq4_counterfactual} shows that both perturbation types consistently degrade accuracy and warning earliness.}

\vspace{+0.2em}
\subsubsection{Concept Sanity Checks.}
\textbf{Shuffled} permutes descriptions within the concept library, whereas \textbf{Random} replaces them with unrelated phrases. The former retains the vocabulary but assigns the wrong meaning to each channel; the latter removes driving relevance altogether. Both preserve the bottleneck structure and dimensionality while breaking semantic identity, yet both degrade all evaluation metrics. The larger decline under Random further indicates that generic language cues cannot replace the task-specific priors distilled from accident reports. These controls rule out the possibility that CARA benefits merely from adding an intermediate representation. \textbf{Instead, its gains depend on concepts retaining risk-relevant meaning throughout prediction.}

\vspace{+0.2em}
\subsubsection{Temporal Necessity Checks.}
These complementary interventions separate short-term semantic stabilization from longer-term evidence accumulation.
Removing \textbf{EMA} exposes the model to noisy frame-level CLIP similarities and weakens the persistence of semantic evidence across adjacent frames. Removing \textbf{TRA}, in contrast, prevents weak but consistent early cues from accumulating into a coherent risk trajectory. The two mechanisms are therefore complementary: EMA stabilizes local concept evidence, whereas TRA models its longer-term risk progression. Both variants degrade accuracy and warning earliness, showing that isolated concept snapshots are insufficient for anticipation. \textbf{CARA instead relies on temporally stabilized activations and their subsequent aggregation to distinguish persistent risk from transient noise.}

Together, the semantic and temporal interventions test complementary failure modes: the former disrupt what each channel means, whereas the latter disrupt how evidence persists and accumulates. \textbf{Their consistent degradation shows that CARA relies on both correct semantic identity and temporal organization.}

\section{Conclusion}

We propose \textbf{CARA}, an intrinsically interpretable framework for collision anticipation via concept-aware risk attention, addressing a key transparency gap in autonomous driving AI. Rather than treating explanation as a post-hoc add-on, CARA formulates anticipation as \emph{concept-mediated sequential reasoning}, in which domain-grounded semantic risk factors are tracked as dynamic intermediate evidence that directly guides spatio-temporal prediction. By deriving risk concepts from accident narratives and integrating their temporally evolving trajectories into the predictive pathway, CARA enables transparent and context-aware risk reasoning without manual concept annotation. Experiments on multiple benchmarks show that CARA consistently improves predictive accuracy and early warning while providing concept-grounded interpretability. Overall, embedding semantic risk evidence directly into the predictive pathway offers a practical direction for safety-critical autonomous driving systems requiring both strong performance and intrinsic interpretability. A remaining limitation is that concept quality still depends on the coverage of the text-derived concept bank. Extending CARA to richer multimodal risk cues is a promising direction for future work.

\begin{acks}
This work was supported by the National Natural Science Foundation of China (NSFC) under Grant No. 62572308.
\end{acks}

\clearpage
\bibliographystyle{ACM-Reference-Format}
\bibliography{MM/main/reference}

\end{document}